\documentclass[twocolumn, showpacs, preprintnumbers, pra, floatfix]{revtex4-1}

\usepackage{graphicx}
\usepackage{epsfig}
\usepackage{bm}
\usepackage{amssymb, amsmath, chemarr}
\usepackage{verbatim}

\def\4he{$^4$He}
\def\3he{$^3$He}
\def\cm3{cm$^{-3}$}

\begin{document}

\title{Vibrational quenching of the electronic ground state in ThO in cold collisions with \3he}

\author{Yat Shan Au,$^{1, 3}$ Colin B. Connolly,$^{1, 3}$ Wolfgang~Ketterle,$^{2, 3}$ and John M. Doyle$^{1, 3}$}
\affiliation{
$^1$Department of Physics, Harvard University, Cambridge, Massachusetts 02138\\
$^2$Department of Physics, Massachusetts Institute of Technology, Cambridge, Massachusetts 02139\\
$^3$Harvard-MIT Center for Ultracold Atoms, Cambridge, Massachusetts 02138
}
\date{\today}

\begin{abstract}
We measure the ratio $\gamma$ of the momentum-transfer to the vibrational quenching cross section for the  X ($^1\Sigma^+$), $\nu=1$, $\mathrm{J=0}$ state of molecular thorium monoxide (ThO) in collisions with atomic \3he between 800~mK and 2.4~K.  We observe indirect evidence for ThO--He van der Waals' complex formation, which has been predicted by theory \cite{Tscherbul2011}.  We determine the 3-body recombination rate constant $\Gamma_3$ at 2.4~K, and establish that the binding energy E$_b >$ 4~K.
\end{abstract}

\pacs{33.50.Hv, 34.50.-s, 37.10.De}


\maketitle

\section*{Introduction}

Cold and ultracold molecules have the potential to revolutionize the fields of precision measurement \cite{Chin2009}, quantum simulation \cite{Ortner2009} and cold controlled chemistry \cite{Krems2008}.  Recent advances in the production of ultracold KRb from laser-cooled atoms had led to the demonstration of quantum state dependent chemical reactions \cite{Ospelkaus2010} and dipolar interactions \cite{Yan2013}.  Cold molecular beams at a few kelvins have promise to drastically improve the sensitivity of electron EDM searches \cite{Hudson2011, Vutha2011b} and enables the differentiation of chiral molecules \cite{Patterson2013a}.

The full potential of molecules, is achieved by exploiting their additional degrees of freedom (DOF), i.e. rotational and vibrational states.  Rotational states could efficiently couple to solid-state devices via microwave radiation \cite{Andre2006};  vibrational states could not only provide sensitive tests for variation in the proton-to-electron mass ratio \cite{Kajita2012}, but also could be used to encode quantum information \cite{Tesch2002}.  Understanding the effect of collisions on these DOF in the molecule will be critical to experimental implementations. 

Compared to the field of atomic collisions \cite{Weiner1999}, the field of cold molecular collisions is still in its infancy.  Despite recent progress in molecule-molecule collision experiments \cite{Sawyer2011, Kirste2012, Stuhl2012}, molecule-atom collisions remain a benchmark platform for understanding of these new systems, particularly in the case of rotational and vibrational state-changing processes.  Photo association \cite{Ulmanis2012}, Stark deceleration \cite{Scharfenberg2010, Fitch2012} and buffer-gas cooling \cite{Bodo2006} are the main experimental tools to study such collisions.  Photo association enables study at ultracold temperatures, but is currently limited to molecules formed by alkali and alkaline earth atoms \cite{Ulmanis2012}; Stark deceleration allows tunable collisional energy with $\sim1$~K resolution, but so far can probe only energies above $\sim30$~K \cite{Bodo2006};  Buffer gas cooling provides access to $\sim 0.5$~K.

In the case of vibrational quenching of a diatomic molecule, there are three temperature ranges: Wigner, van der Waals (vdW), and high temperature.  At intermediate temperatures, comparable to the energy of the vdW interaction, the potential can support shape and Feshbach resonances, resulting generally in an increased quenching rate at lower temperature \cite{Quemener2010}.  Therefore, study below the vdW temperature ($\sim$ 10~K) can provide useful constraints on potential energy surfaces (PESs). 

In this paper, we focus on vibrational quenching collisions between molecular thorium monoxide (ThO) and atomic \3he in the intermediate vdW regime.  We determine that vibrational states relax faster than vdW molecule formation.  Our work also extends the experimental study of vibrational quenching to include closed-shell molecules.

\section*{Experimental Setup}

At the heart of our work is the production of a cold ThO molecular gas in the presence of cold \3he.  We produce cold ThO molecules using the technique of buffer gas cooling \cite{Doyle1999}.  Our apparatus is similar to that of Ref.~\cite{Johnson2010}.  Using a dilution refrigerator, we cool a copper cell to a temperature of 0.8--4~K.  We ablate a ThO$_2$ ceramic target with a YAG laser pulse of a few mJ to produce ThO molecules, which then collide with \3he and thermalize their translational motion to the cell's temperature.  To calibration the density of \3he, we simultaneously ablate chromium (Cr) metal to produce a gas of cold atomic Cr and measure the Cr diffusion rate in the buffer gas.  Cr is detected via laser absorption spectroscopy using the 3d$^5$(6S) 4s ($a$ $^7$S$_3$) $\rightarrow$ 3d$^5$(6S) 4p ($z$ $^7$P$_ 4$) transition at 426 nm.  We adapt the value of $\sigma_{d} = 1.1 \times 10^{-14}$~cm$^2$ for the Cr-$^3$He diffusion cross section \cite{Weinstein2001}.  As $\sigma_{d}$ can potentially vary with temperature, as observed in other atomic \cite{Lu2008, Tscherbul2009} and molecular systems \cite{Lu2009}, we account for this potential effect as a systematic error. 

\section*{van der Waals molecules}

\begin{figure}
\includegraphics[width=8.6 cm]{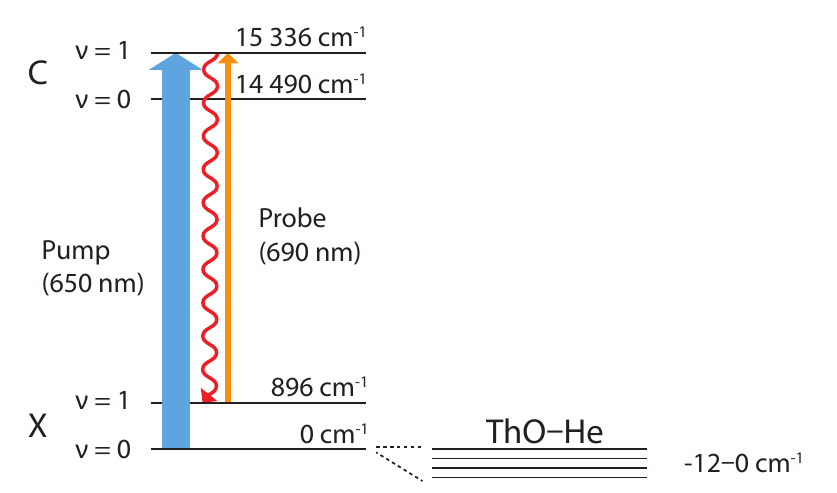}
\caption{\label{fig:level} (Color online)  Relevant energy levels of the ThO molecule.  The letters refer to the electronic states, X ($^1\Sigma^+$) and C (77\% $^1\Pi$, 20\% $^3\Pi$), respectively.  All measurements were performed in the ground rotational levels, $\mathrm{J=0}$ for the X state and $\mathrm{J=1}$ for the C state.  Optical pumping via the X ($\nu=0$) $\rightarrow$ C ($\nu=1$) R(0) transition at 650~nm is used to transfer population to the X ($\nu=1$) state, which is probed using the X ($\nu=1$) $\rightarrow$ C ($\nu=1$) R(0) transition at 690~nm.}
\end{figure}

It has been predicted theoretically \cite{Tscherbul2011} that ThO forms ThO-He  van der Waals (vdW) molecules in the presence of dense helium gas below 10~K (Fig.~\ref{fig:level}).  Noble gas vdW  molecules  have been studied in a wide range of experiments, from supersonic jets \cite{Koperski2003}, bulk liquid helium \cite{Persson1996} and buffer gas cells \cite{Brahms2011}.  In our setup, vdW can be formed via a 3-body recombination process.

\begin{equation}
\mathrm{ThO} + \mathrm{He}  + \mathrm{He} \xrightleftharpoons[\Gamma_2]{\Gamma_3} \mathrm{ThO-He} + \mathrm{He}
\end{equation}

\noindent where $\Gamma_2$ and $\Gamma_3$ are rate coefficients for 2-body dissociation and 3-body combination, respectively.  Suppose we begin our experiment by injecting free ThO molecules into a gas of helium.  The vdW molecule formation dynamics of the system can be described by:

\begin{align}
\frac{d\mathrm{[ThO]}}{dt} &= -\frac{\Gamma_D}{\mathrm{[He]}}\mathrm{[ThO]}
+\Gamma_2\mathrm{[He]}\mathrm{[ThO-He]}  \nonumber\\
& \quad -\Gamma_3\mathrm{[He]^2}\mathrm{[ThO]}
\label{eq:vdw_ThO}
\end{align}

\begin{align}
\frac{d\mathrm{[ThO-He]}}{dt} &= -\frac{\Gamma_D}{\mathrm{[He]}}\mathrm{[ThO-He]}
-\Gamma_2\mathrm{[He]}\mathrm{[ThO-He]} \nonumber\\
& \quad +\Gamma_3\mathrm{[He]^2}\mathrm{[ThO]} 
\label{eq:vdw_ThO_He}
\end{align}

\noindent where $\Gamma_D$ is the diffusive rate constant.  Note that the diffusive decay rate is inversely proportional to the helium density.

\begin{figure}
\includegraphics[width=8.6 cm]{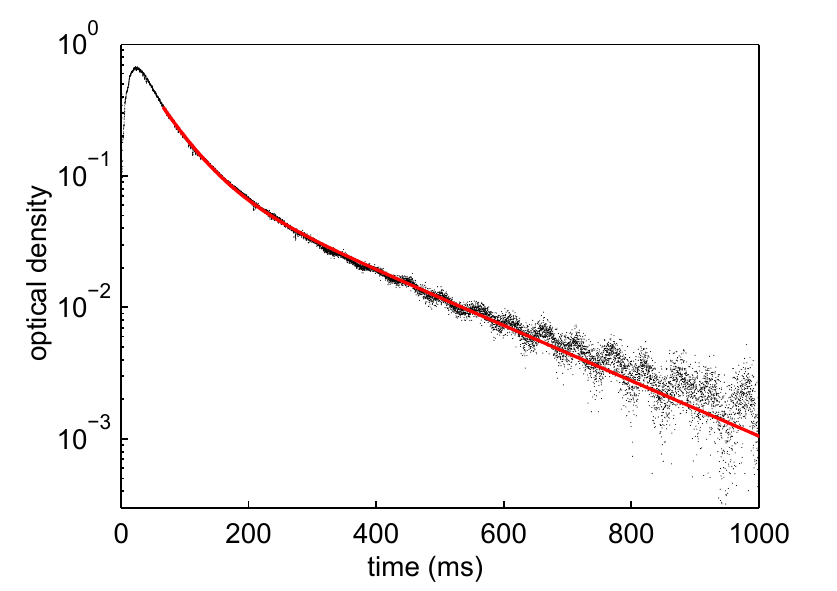}
\caption{\label{fig:ThO_vdw_data}  (Color online)  The optical density (OD) decay of ThO in its X ($\nu=1$, $\mathrm{J=0}$) state at high helium density at 2.4~K.  Solid (red) line is fit to a sum of two exponentials.  We verify that both diffusion modes and temperature stabilize within the first 50~ms by observing decay of atomic chromium produced simultaneously (not shown).  The apparent oscillation visible at late time (small OD) is caused by mechanical vibration of the apparatus, and it has insignificant effect on the early-time (large OD) fit.}
\end{figure}

The optical density ($\propto$ molecule number) of free ThO molecules evolves as a sum of two exponential decays (see Fig.~\ref{fig:ThO_vdw_data}) with time constants given by:

\begin{eqnarray}
\tau_s & = & {\left(\frac{\Gamma_D}{\mathrm{[He]}} +  \Gamma_2 \mathrm{[He]} +  \Gamma_3{\mathrm{[He]}}^2\right)}^{-1} \label{eq:time_short}\\
\tau_l & = & \frac{\mathrm{[He]}}{\Gamma_D} \label{eq:time_long}
\end{eqnarray}

\begin{figure}
\includegraphics[width=8.6 cm]{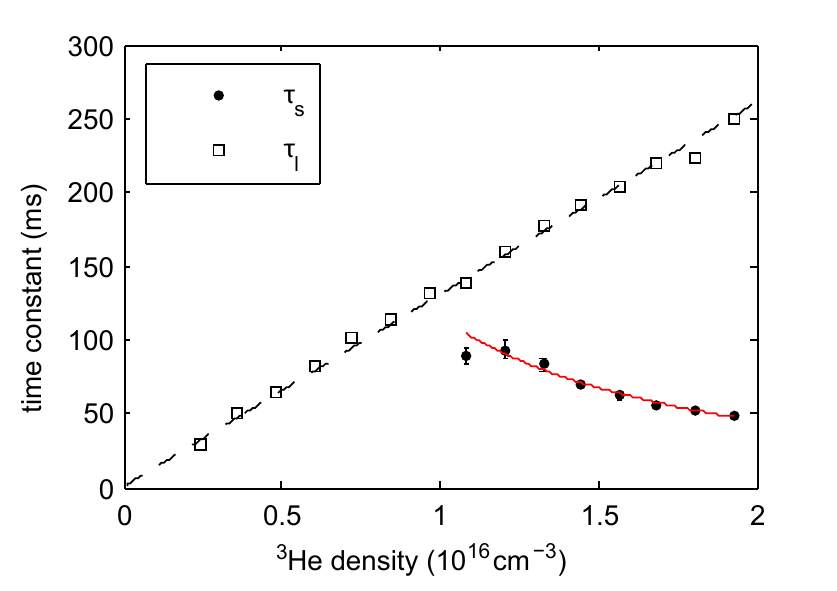}
\caption{\label{fig:ThO_vdw_fit}  (Color online)  Fits to decay time constants $\tau_s$ and $\tau_l$ in Eq.~\ref{eq:time_short} and Eq.~\ref{eq:time_long} at various density of helium at 2.4~K.  Open squares (solid dot with error bars) denote fitted value for $\tau_l$ ($\tau_s$).  Error bars are statistical.  Dashed line (solid red line) are fits to Eq.~\ref{eq:time_long} (Eq.~\ref{eq:time_short}), where the diffusive rate constant $\Gamma_D$ (3-body recombination rate $\Gamma_3$ and 2-body decay rate $\Gamma_2$) in Eq.~\ref{eq:vdw_ThO} is extracted.}
\end{figure}

Here $\tau_s$ and $\tau_l$ represent time constants for the return to thermal equilibrium between ThO and ThO-He and for diffusive decay, respectively.  By repeating the experiment at various helium densities (Fig.~\ref{fig:ThO_vdw_fit}), we extract these time constants.  The linear dependence of $\tau_l$ on helium density confirms our interpretation of it as the diffusive decay time constant.   From this data, we determine the ratio between $\sigma_d$ for ThO-$^3$He and Cr-$^3$He collisions to be $1.75 \pm 0.02$ at 2.4~K.  By fitting the $\tau_s$ data to Eq.~\ref{eq:time_short} (Fig.~\ref{fig:ThO_vdw_fit}), we also determine the 3-body recombination rate constant to be $\Gamma_3 = 2.5 \pm 0.7 \times 10^{-32}$~cm$^6$s$^{-1}$, which is comparable to the value of $10^{-31}$~cm$^6$s$^{-1}$ measured for the Ag-$^3$He system \cite{Brahms2011}.  Although we extract a value for $\Gamma_2$ from the fit, its interpretation is not straightforward.  In Eq.~\ref{eq:vdw_ThO} and Eq.~\ref{eq:vdw_ThO_He}, we simplify our system to possess only one ThO--He vdW bound state.  Under such simplification, $\Gamma_2 = c \cdot e^{-kT/E_{b}}$, for some constant $c$ and binding energy $E_{b}$, both of which can be determined by repeating the experiment at a different temperature.  

From Ref.~\cite{Tscherbul2011}, we estimate there are about 30  ThO-$^3$He vdW bound states.  No specific prediction was made in Ref.~\cite{Tscherbul2011} for this system.  A similar number of vdW states has been observed in other atom-$^3$He and  atom-$^4$He systems \cite{Brahms2011}.   As a result,

\begin{equation}
\Gamma_2 = \sum_i c_i \cdot f_i \cdot e^{-kT/E_{b,i}}
\end{equation}

\noindent where $f_i$ is the fractional occupation for bound state $i$.  Unlike the measured value of $\Gamma_3$, which starts from an electronic and vibrational state, the measured value of $\Gamma_2$ is a thermal averaged rate convoluted with the formation ($f_i$) dynamics and vdW bound state-changing collisions.  The highest temperature at which we observed a decay of the form predicted by Eq.~\ref{eq:time_short} and Eq.~\ref{eq:time_long} is 4~K, and thus we conclude there exist vdW bound states with binding energy E$_b > 4$~K.

\section*{Vibrational quenching}

To measure vibrational quenching in ThO molecules, we use an optical pump-and-probe technique.  We transfer molecule population from the absolute ground state to the first-vibrationally-excited state via the X ($\nu = 0$) $\rightarrow$ C ($\nu = 1$) R(0) transition at 650~nm.  We probe the X ($\nu = 1$) state using the X ($\nu = 0$) $\rightarrow$ C ($\nu = 1$) R(0) transition at 690~nm (Fig.~\ref{fig:level}).  All measurements were performed in the ground rotational levels, $\mathrm{J=0}$ for the X state and $\mathrm{J=1}$ for the C state.

\begin{figure}
\includegraphics[width=8.6 cm]{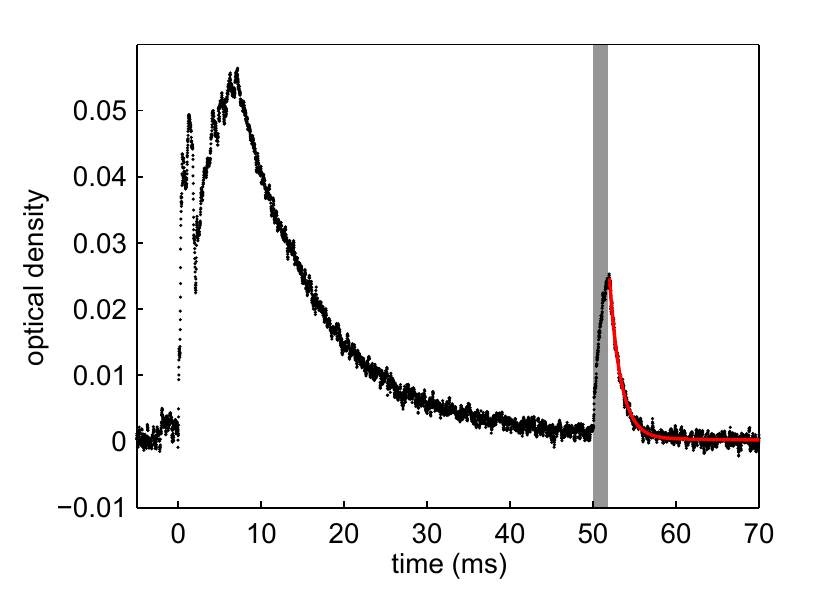}
\caption{\label{fig:ThO_v1_data}  (Color online)  The optical density (OD) of ThO in its X ($\nu=1$, $\mathrm{J=0}$) state at 1.2~K.  The initial OD, produced by molecule production via laser ablation, exhibits strong shot-to-shot fluctuation in both magnitude and duration, and therefore, it is unsuitable for the study of vibrational quenching.  At 50~ms, a short optical pumping pulse (shaded area) was used to transfer population from X ($v=0$, $\mathrm{J=0}$) state.  Solid (red) line is fit to a single exponential decay.}
\end{figure}

As shown in Fig.~\ref{fig:ThO_v1_data}, laser ablation of the solid ThO$_2$ target produces significant population in the X ($\nu = 1$) state.  However, such population fluctuates strongly in both magnitude and duration from shot-to-shot, presumably due to different initial populations of higher-lying states, which then decay to the $\nu=0$ state through the $\nu=1$ state. Hence, we cannot use the initial state population to study the dynamic of vibrational quenching.

Instead, we wait for the initial population to decay and  then apply an optical pumping pulse 50~ms after ablation to transfer population from the $\nu=0$ to the $\nu=1$ state.  We fit a single exponential decay to the $\nu=1$ population created by optical pumping, and repeat the experiment at over range of helium densities.  We obtain the same results when the optical pumping pulse is applied 10~ms later instead, and therefore conclude that the contribution to the fitted decay from the initial population is insignificant.

\begin{figure}
\includegraphics[width=8.6 cm]{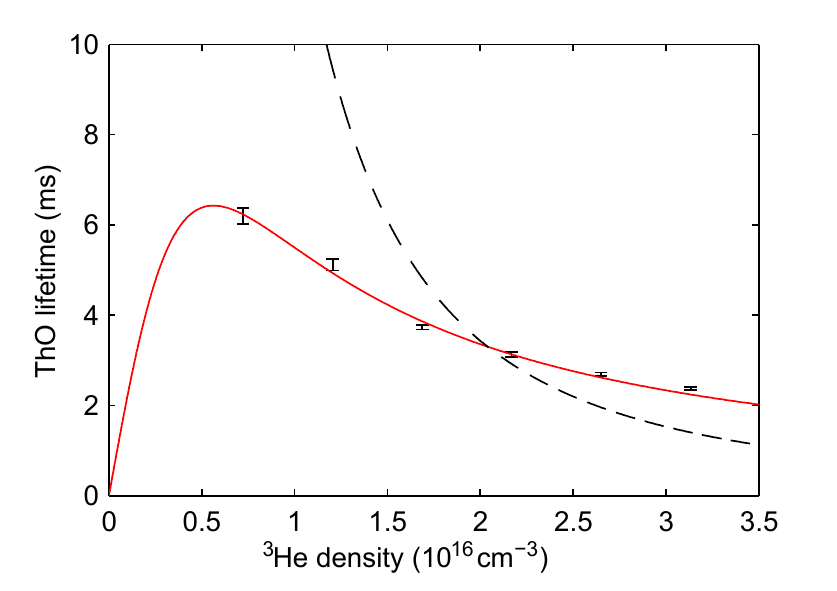}
\caption{\label{fig:ThO_v1_fit}  (Color online)  Observed vibrational quenching of the X ($\nu=1$, $\mathrm{J=0}$) state of ThO at 1.2~K.  Error bars are statistical uncertainties. \3he density is determined from the diffusive decay of Cr produced simultaneously \cite{Weinstein2001}.  The solid red line is a combined diffusion, 2-body and 3-body fit to data (Eq.~\ref{Eq:v_1_fit}).  The dashed line is a 3-body fit to all the data.  The effect of diffusion is insignificant at $n_{\mathrm{He}} \gtrsim 1.2 \times 10^{16}$~cm$^{-3}$. }
\end{figure}

The measured ThO X ($\nu = 1$) state lifetime $\tau_\nu$ at various helium densities $n_{\mathrm{He}}$ is shown in Fig.~\ref{fig:ThO_v1_fit}.  There are three mechanisms for decay from this state: diffusion ($\tau_d \propto n_{\mathrm{He}}$), vibrational quenching ($\tau_q \propto n_{\mathrm{He}}^{-1}$) and vdW molecule formation ($\tau_3 \propto n_{\mathrm{He}}^{-2}$).

\begin{eqnarray}
\tau_{\nu=1} &=& \left(\frac{1}{\tau_d} + \frac{1}{\tau_\nu} + \frac{1}{\tau_3}\right)^{-1} \nonumber \\ &=&\left(\frac{\Gamma_d}{n_{\mathrm{He}}} + \Gamma_\nu n_{\mathrm{He}} + \Gamma_3 n_{{\mathrm{He}}}^{2}\right)^{-1}
\label{Eq:v_1_fit}
\end{eqnarray}
\noindent for some rate constants $\Gamma$'s.

At $n_{\mathrm{He}} \gtrsim 1.2 \times 10^{16}$~cm$^{-3}$, the effect of diffusion is insignificant.  No theoretical prediction has been made for vdW molecule formation in the X ($\nu = 1$) state.  The combined fit indicates that 3-body recombination accounts for less than 25\% of the observed $\nu=1$ decay.

\begin{figure}
\includegraphics[width=8.6cm]{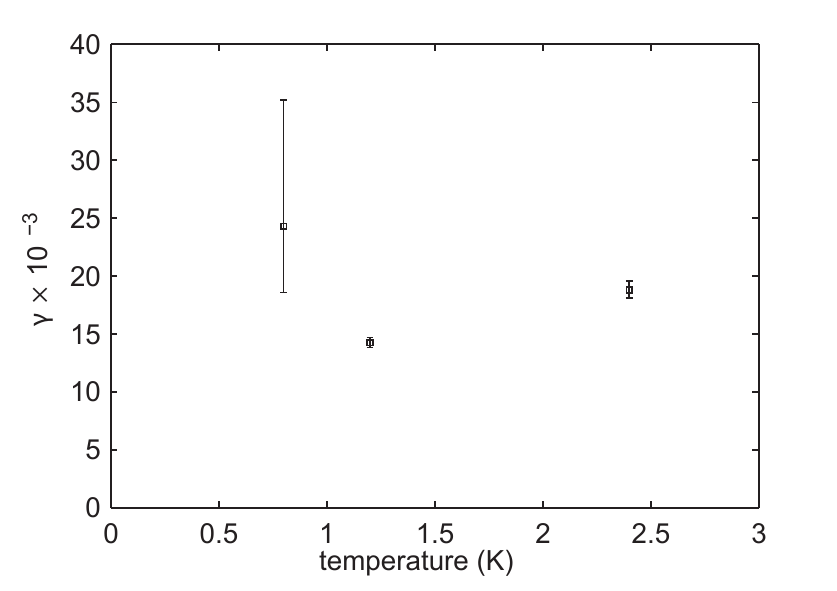}
\caption{\label{fig:ThO_v1_results}  Collisional quenching of the first vibrationally excited state of ThO in collisions with \3he.  Shown is the ratio of the momentum-transfer to the vibrational quenching cross section for the  X ($^1\Sigma^+$), $\nu=1$, $\mathrm{J=0}$ state versus temperature.}
\end{figure}

We measure vibrational quenching at three temperatures (Fig.~\ref{fig:ThO_v1_results}).  The relatively large error bar for the 800~mK measurement is caused by small signals, probably due to the lack of free ground state molecules at lower temperature.

Although we can in principle transfer population to an initial $\mathrm{J > 0}$ state, mixing of rotational states in collisions is expected to be more rapid than vibrational quenching \cite{Tscherbul2011}.  Therefore, our experiment is confined to the $\mathrm{J=0}$ state.

Compared to the measured vibrational quenching rates in CaH-He and CO-He, our value for ThO-$^3$He is a few orders of magnitude faster.  One possible reason is that more shape and Feshbach resonances are supported by a stronger vdW interaction between ThO and $^3$He.  The results remain to be explained by detailed theoretical calculations.

\section*{Conclusion}
We obtain indirect evidence for van der Waals' bound state between ThO and \3he by observing double-exponential decay at high helium density and low temperature.  We determine the 3-body recombination rate constant $\Gamma_3 = 2.5 \pm 0.7 \times 10^{-32}$~cm$^6$s$^{-1}$ at 2.4~K.

We also determined the ratio $\gamma$ of the momentum-transfer to the vibrational quenching cross section of the  X ($^1\Sigma^+$), $\nu=1$, $\mathrm{J=0}$ state.  We find $\gamma \sim 10^4$, which suggests sympathetic cooling of the vibrational modes can be slow, but the long collisional lifetime of the $\nu =1$ state can also lead to a wide range of applications such as vibrational qubits \cite{Tesch2002}.

\begin{acknowledgments}

We acknowledge Timur Tscherbul for helpful theoretical inputs and Elizabeth Petrik for assistance in target preparation.  This work was supported by the NSF through the Harvard-MIT Center for Ultracold Atoms.

\end{acknowledgments}

\bibliographystyle{apsrev}
\bibliography{ThO_PRA_02}

\end{document}